\documentclass[reprint,prb,showpacs,superscriptaddress, english]{revtex4-1}
\usepackage[utf8]{inputenc}
\usepackage{color}
\usepackage{amsmath}
\usepackage{amssymb}
\usepackage{graphicx}
\usepackage{esint}
\usepackage[unicode=true,
 bookmarks=false,
 breaklinks=true,pdfborder={0 0 0},backref=false,colorlinks=true]
 {hyperref}
\usepackage{breakurl}

\makeatletter

\@ifundefined{textcolor}{}
{%
 \definecolor{BLACK}{gray}{0}
 \definecolor{WHITE}{gray}{1}
 \definecolor{RED}{rgb}{1,0,0}
 \definecolor{GREEN}{rgb}{0,1,0}
 \definecolor{BLUE}{rgb}{0,0,1}
 \definecolor{CYAN}{cmyk}{1,0,0,0}
 \definecolor{MAGENTA}{cmyk}{0,1,0,0}
 \definecolor{YELLOW}{cmyk}{0,0,1,0}
}

\usepackage{graphicx}
\usepackage{epstopdf}\usepackage{subfigure}
\usepackage{float}
\usepackage{amsfonts}
\usepackage{bm}
\usepackage{subfigure}
\usepackage{caption}

\makeatother

\begin{document}

\title{Instability of Dirac semimetal phase under strong magnetic field}

\author{Zhida Song}
\affiliation{Beijing National Laboratory for Condensed Matter Physics and Institute
of Physics, Chinese Academy of Sciences, Beijing 100190, China}
\affiliation{University of Chinese Academy of Sciences, Beijing 100049, China}

\author{Zhong Fang}
\affiliation{Beijing National Laboratory for Condensed Matter Physics and Institute
of Physics, Chinese Academy of Sciences, Beijing 100190, China}
\affiliation{Collaborative Innovation Center of Quantum Matter, Beijing, 100084,
China}

\author{Xi Dai}
\email{daix@aphy.iphy.ac.cn}
\affiliation{Beijing National Laboratory for Condensed Matter Physics and Institute
of Physics, Chinese Academy of Sciences, Beijing 100190, China}
\affiliation{Collaborative Innovation Center of Quantum Matter, Beijing, 100084,
China}

\begin{abstract}
The quantum limit can be easily reached in the Dirac semimetals under the magnetic field,
 which will lead to some exotic many-body physics due to the high degeneracy of the
 topological zeroth Landau bands (LBs).
By solving the effective Hamiltonian, which is derived by tracing out
 the high energy degrees of freedom, at the self-consistent  mean field level,
 we have systematically studied the instability of Dirac semimetal under a strong magnetic field.
A charge density wave (CDW) phase and a polarized nematic phase formed by
 “exciton condensation” are predicted as the ground state for the tilted and untilted bands, respectively.
Furthermore, we propose that, distinguished from the CDW phase, the nematic phase can be identified in experiments by anisotropic transport and Raman scattering.
\end{abstract}
\maketitle

\section{Introduction}

Searching for new states of matter in solid materials is one of the
key problems in condensed matter physics, which attracts lots of research
interests recently. External magnetic field has provided an additional
dimension for such studies, leading to surprisingly rich phenomenons
and phases in two dimensional electron gas systems already, e.g. the
integer \cite{IQHE} and fractional \cite{FQHE} quantum Hall effects, the Wigner crystal phase
as well as the nematic phases \cite{PhysRevLett.83.824,Du1999389,Feldman316}.
The high degeneracy of Landau levels
resulted from the Landau quantization of the electronic wave functions
is the main origin of the instability towards the various exotic
phases mentioned above. In three dimensional systems, the Landau quantization
only happens in the plane perpendicular to the magnetic field and the energy dispersion along the field direction remains unchanged.
For ordinary semiconductor system with quadratic band dispersion,
the high degeneracy of the LBs leads to almost perfect nesting
of the “Fermi surfaces” along the field direction, as illustrated
schematically in Fig. (\ref{fig:channel}).
Such a nesting effect will be greatly enhanced in the so called quantum limit,
 where only the lowest LB cuts through the Fermi level, and the field induced
 symmetry breaking phases such as CDW \cite{celli_ground_1965,PhysRevB.30.7009,alicea_bismuth_2009},
 spin-density wave \cite{celli_ground_1965,noauthor_possible_1987,tesanovic_multivalley_1987}, valley-density wave \cite{tesanovic_multivalley_1987},
 will be stabilized as the ground state.

It is very difficult to reach the quantum limit in normal semiconductors
and semimetals and the experimental observation of the field induced
CDW phase in  real materials, i.e. Bi and Sb, is still under debate. \cite{Behnia1729,Li547}
The recently discovered topological semimetals provide a new platform
for the search of new exotic phenomenons under magnetic field.
 \cite{nielsen_adler-bell-jackiw_1983, son_chiral_2013, burkov_chiral_2014,
  parameswaran_probing_2014, xiong_evidence_2015, lu_weak_2015, song_detecting_2016,
  rinkel2016signatures, zhang_magnetic_2015}
For Dirac \cite{young_dirac_2012, wang_dirac_2012,
 yang_classification_2014}
 or Weyl \cite{burkov_weyl_2011, wan_topological_2011, weng_weyl_2015,
 huang_weyl_2015, lv_observation_2015, xu_discovery_2015,nie_topological_2017} semimetals,
 where the Fermi level is very close to the
Dirac or Weyl points, the quantum limit can be easily reached even under
a weak magnetic field and more fruitful many-body physics can be
realized due to the extra valley and orbital degrees of freedom.
\cite{miransky_quantum_2015,shovkovy_magnetic_2013}
For instance, in strong magnetic field, the Weyl semimetal is found to be
 stabilized as a chiral-symmetry-breaking CDW state. \cite{Weyl_ACDW,Xiaotian_CDW}
In the present Letter, we systematically study the possible instabilities
of Dirac semimetal state under the magnetic field in the quantum
limit.
We find that, besides the CDW phase, a new state,
 the polarized nematic phase can be stabilized in a large part of the phase diagram.
Such an exotic phase is caused by the ``exciton condensation" between
 the two zeroth LBs, which breaks both the rotational symmetry
 and the inversion symmetry, leading to a number
 of important physical consequences in transport and optical experiments.

\section{Model}

The Dirac semimetals can be divided into two categories by whether
the Dirac points (DPs) are located on high symmetry lines or points \cite{yang_classification_2014}
of the Brillouin zone (BZ).
In this Letter, we will focus on the first category, where the DPs are protected
by the crystalline symmetry along the high symmetry lines and always appear in pairs due to the presence
of the time reversal symmetry. The typical example of such type of
materials is $\mathrm{Na_{3}Bi}$ \cite{wang_dirac_2012}, where the
DPs are generated by the crossings of two doubly degenerate bands
along the $z$ axis.
The low energy physics of such type of Dirac semimetal can be well described
by the following k$\cdot$p model,
\begin{equation}
H^0= C(k_z) +
\begin{pmatrix}M\left(k_{z}\right) & -v\hbar k_{-} & \gamma\left(\mathbf{k}\right) & 0\\
-v\hbar k_{+} & -M\left(k_{z}\right) & 0 & \gamma\left(\mathbf{k}\right)\\
\gamma^{*}\left(\mathbf{k}\right) & 0 & -M\left(k_{z}\right) & v\hbar k_{-}\\
0 & \gamma^{*}\left(\mathbf{k}\right) & v\hbar k_{+} & M\left(k_{z}\right)
\end{pmatrix}\label{eq:H0mat}
\end{equation}
Here $C(k_z) = C_0(\cos a_0 k_z - \cos a_0 k_c)$,
 $ M\left(k_{z}\right)=M_{0}\left(\cos a_{0}k_{z}-\cos a_{0}k_{c}\right)$,
  $k_{\pm}=k_{x}\pm ik_{y}$, $v$ is the velocity in $xy$ plane,
 $a_{0}$ is the lattice
 along $k_{z}$, and $\pm k_{c}$ are the locations of DPs.
The bases of the k$\cdot$p model can be labeled by their main orbital characters as
$\left|P\frac{3}{2}\right\rangle $, $\left|S\frac{1}{2}\right\rangle $,
$\left|S-\frac{1}{2}\right\rangle $, $\left|P-\frac{3}{2}\right\rangle $,
respectively.
The first term in Eq. (\ref{eq:H0mat}) plays an important role in the formation of
 type II Weyl points. \cite{Weyl_type2,XuYong_TypeII}
While as long as $|C_0|<|M_0|$, that is the case we focus on,
 the  $C(k_z)$  term will just tilt the DPs
 and change the ellipsoidal Fermi surface to a pyriform one.
Even so, as shown in the following, this term will play an important role in determining
 whether the CDW or nematic phase will be stabilized.
The high order term $\gamma(\mathbf{k})$ won't play any important role for the physics discussed here
 and so will be neglected in the rest of the Letter.

The external magnetic field $B$ is applied along the $z$ direction.
Adopt the Landau gauge $\mathbf{A}=\left(-yB,0,0\right)$,
 which leaves $k_x$ and $k_z$ still good quantum numbers,
 the LB eigenenergies and eigenstates can be solved analytically (see appendix \ref{sec:solution}).
As shown in Fig. (\ref{fig:channel}), the two zeroth LBs
 disperse linearly and cross with each other at the DPs.
The quantum limit can be reached by increasing the magnetic field such that only the zeroth
 LBs cuts through the Fermi level.
In the present work we are only interested in the instability
 in the quantum limit, therefore we only keep the zeroth LBs in the non-interacting Hamiltonian
\begin{equation}
\hat{H}^{0}= \sum_{a k_x k_z} \epsilon_{k_{z}a} \hat{\psi}_{k_{x}k_{z}a}^{\dagger} \hat{\psi}_{k_{x}k_{z}a}\label{eq:H0}
\end{equation}
\begin{equation}
\epsilon_{k_{z}a}=
\begin{cases}
C(k_z)-M\left(k_{z}\right) & a=c\\
C(k_z)+M\left(k_{z}\right) & a=v
\end{cases}
\end{equation}
Here $c$ and $v$ represent the conduction band (red band in Fig. (\ref{fig:channel}))
 and valence band (blue band in Fig. (\ref{fig:channel})),
 which are formed by $\left|S\frac{1}{2}\right\rangle $ and $\left|P-\frac{3}{2}\right\rangle $
 states respectively.
Since they belong to different eigenvalues of $C_{6}$, the crossings at $\pm k_{c}$ are protected by rotational symmetry
 and will persist even if  nonzero $\gamma\left(\mathbf{k}\right)$
 presents.

Notice that the Zeeman's coupling between the magnetic field and the field-free orbitals
is neglected here.
In a first principle study of the effective g factor, \cite{gfactor}
 we show that the Zeeman’s splitting in a typical Dirac semimetal
 under magnetic field as strong as 100T is just about 5meV, which is much smaller than the
 band inversion energy ($M_0$) and so would not affect the discussion qualitatively.

\begin{figure}
\begin{centering}
\includegraphics[width=1\linewidth]{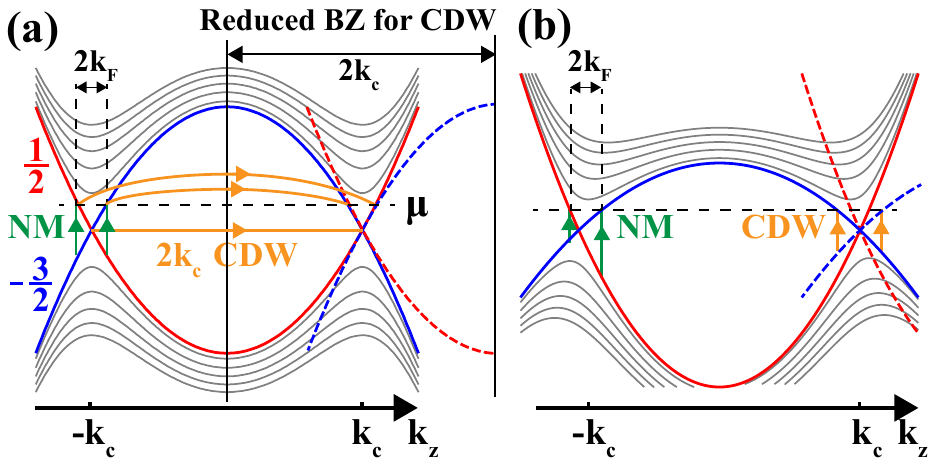}
\par\end{centering}

\protect\caption{\label{fig:channel}
Band structure and instability channels.
The two zeroth LBs, referred as the conductance and valence bands,
 are plotted in red and blue respectively,
 and the high energy LBs are plotted in gray.
(a) shows the nematic phase channel (green
 arrows) and three CDW phase channels (orange arrows) in an untilted
 band structure ($C_0=0$).
 As argued in the text, only the nematic phase and the $2k_c$ CDW phase
 can be realized in low density limit.
 Here the dashed colored lines represent the folded bands in the $2k_c$ CDW phase.
(b) shows the two channels in a tilted band structure ($C_0\neq 0$).
 It is apparent that the kinetic energy cost of the $2k_c$ CDW order
 will be significantly lowered by the tilting.
}
\end{figure}

\section{Effective interaction}

To explore the stability of the above system under Coulomb repulsive interaction,
 we need to derive an effective interaction
 for the zeroth LBs by tracing out all the high LBs.
Take the random phase approximation (RPA), we get
\begin{align}
\hat{H}_{\mathrm{int}}^{\mathrm{eff}} & =\frac{1}{2\Omega}\sum_{\mathbf{q}\neq0}
 \sum_{k_x k_z} \sum_{p_x p_z} \sum_{ab} e^{il_{B}^{2}q_{y} \left(k_{x}-p_{x}\right)} W\left(\mathbf{q}\right)\nonumber \\
 \times & \hat{\psi}_{k_{x}+q_{x,}k_{z}+q_{z},a}^{\dagger} \hat{\psi}_{p_{x},p_{z},b}^{\dagger} \hat{\psi}_{p_{x}+q_{x},p_{z}+q_{z},b} \hat{\psi}_{k_{x},k_{z},a}\label{eq:Hint}
\end{align}
\begin{align}
W\left(\mathbf{q}\right) & =\frac{e^{2}}{\epsilon_{0}\kappa\left(\mathbf{q}\right)\mathbf{q}^{2}}e^{-\frac{1}{2}l_{B}^{2}\mathbf{q}_{\perp}^{2}}
\end{align}
where $\mathbf{q}_{\perp}=\left(q_{x},q_{y}\right)$, $l_B=\sqrt{\frac{\hbar}{eB}}$ is
 the magnetic length, $\kappa(\mathbf{q})$
 is the effective dielectric function, and $\Omega$ is the sample volume.
Details of the RPA derivation and the discussion of the dielectric function
 are given in appendix \ref{sec:RPA}.
As shown below, the long wave part of the interaction
contributes the most in both of the possible instabilities, thus we can approximate $\kappa\left(\mathbf{q}\right)$
 by a dielectric constant
$ \kappa =\kappa_{0} +  \frac{1}{3} \kappa_{z} + \frac{2}{3}\kappa_{xy} $,
where
\begin{equation}
\kappa_{z} \approx \frac{e{}^{2}u}{3\pi^{2} \epsilon_{0}v^{2}\hbar} \left(0.9+\ln\left( \frac{M_{0}l_{B}}{v\hbar} \right) \right) \label{eq:kapp-para}
\end{equation}
\begin{equation}
\kappa_{xy} \approx \frac{e{}^{2}}{4\pi^{2} \epsilon_{0} u\hbar} \left(0.6 + \ln \left( \frac{M_{0}l_{B}}{v\hbar} \right) \right) \label{eq:kapp-perp}
\end{equation}
are the dielectric constants from high LBs,
 $\kappa_{0}$ is the dielectric constant from the core electrons,
 and $u=\frac{1}{\hbar}M_{0}a_{0}\sin\left(a_{0}k_{c}\right)$ is the
 Dirac velocity along $z$ direction.
The derivation of such dielectric constants is given in appendix \ref{sec:kap}.
It should be aware that the results
 in Eq. (\ref{eq:kapp-para})-(\ref{eq:kapp-perp}) are not only applied to this particular model,
in fact it is universal for all the Dirac/Weyl semimetals. One of
the important features for the above effective interaction is that
its strength can be tuned by external magnetic field, which is a bit unusual
 in condensed matter physics. The mechanism is
easy to be understood, that is, the energy gap between the zeroth and high
LBs increases with the field strength, which weakens the
screening effect.

\section{COHSEX method}

\begin{figure}
\begin{centering}
\includegraphics[width=1\linewidth]{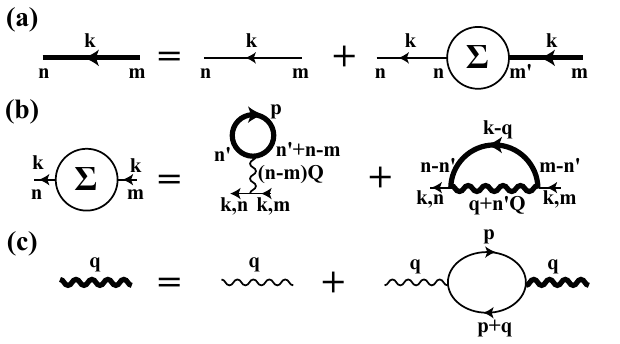}
\par\end{centering}

\protect\caption{\label{fig:COHSEX}Feynman diagrams for the COHSEX method.
The full and free Green's functions are represented by thick and thin lines,
 respectively.
Correspondingly, the screened effective interaction $W^\mathrm{S}$ and the bare
 effective interaction $W$ are represented by thick and thin wavy lines,
 respectively.
(a) is the diagram for the Dyson's equation.
(b) is the diagram for the self energy.
(c) is the diagram for the screened effective interaction, where the Green's functions
 participating the screening are approximated by the free Green's functions.
}

\end{figure}

It is well known that the direct Hartree-Fock mean field approximation for metals with long
range Coulomb interaction leads to a singular Fermi velocity because
of a logarithmic divergence in the exchange channel.
To handle this problem, we adopt the ``Coulomb hole plus screened exchange''
(COHSEX) method, which is a simplified version of the GW method \cite{hedin_new_1965}.
Applying this method to our model, the  self energy consists of
 a direct Hartree term $\Sigma^\mathrm{H} $
 and a screened exchange term $\Sigma^\mathrm{E}$ where the interaction
 is not only screened by high LB electrons but also the zeroth LB electrons.
As explained in the next section, in low carrier density limit, the
 system has a CDW instability at $Q=2k_c$.
For convenience of calculation, we take the commensurate limit by setting
 $k_{c}=\frac{\pi}{D_{c}a_{0}}$, where $D_{c}$ is an integer.
The BZ will be folded $D_c$ times if the CDW order presents.
Thus, in general, we can define the
 Green's function as
 $G_{an,bm}(k_x,k_z,t)=
 \langle T_{t} \hat{\psi}_{k_x,k_z+nQ,a}(t) \hat{\psi}_{k_x,k_z+mQ,b}^\dagger(0) \rangle$,
 where $n,m=0\cdots D_c-1$ is the sub-BZ index and $k_z$ takes value in the reduced BZ:
 $0\le k_z<Q$.
Then the self energy can be expressed as (Fig. (\ref{fig:COHSEX}b))
\begin{align}
\Sigma_{an,bm}^{\mathrm{H}} \left(k_{x}k_{z}\right) &= \delta_{ab} W \left(0,0,\left(n-m\right)Q\right) \int
 \frac{dp_{x}dp_{z}d\omega}{\left(2\pi\right)^{3}}  \nonumber \\
 \times & \sum_{cn^{\prime}} G_{c,n^{\prime}+n-m; c,n^{\prime}}\left(p_{x},p_{z},\omega\right)e^{i\omega0^{+}} \label{eq:Sigma-H}
\end{align}
\begin{align}
\Sigma_{an,bm}^{\mathrm{E}} \left(k_{x}k_{z}\right) &= -\sum_{n^{\prime}} \int \frac{d^{2}\mathbf{q}_{\perp}}{\left(2\pi\right)^{2}} \int\frac{dp_{z}d\omega}{\left(2\pi\right)^{2}} \nonumber \\ \times&W^{\mathrm{S}} \left(\omega=0,\mathbf{q}_{\perp},k_{z}-p_{z}+n^{\prime}Q\right) \nonumber \\ \times & G_{a,n-n^{\prime};b,m-n^{\prime}}\left(k_{x}-q_x,p_{z},\omega\right)e^{i\omega0^{+}}
\label{eq:Sigma-E}
\end{align}
where $W^\mathrm{S}(\omega=0)$ is the \textit{static} screened interaction.
Here we approximate the Green's function screening $W^\mathrm{S}$
 by the free Green's function at zero doping, as shown in Fig. (\ref{fig:COHSEX}c).
Such an approximated screened interaction can be derived analytically
\begin{equation}
W^{\mathrm{S}}\left(\omega=0,\mathbf{q}\right)  = \frac{e^{2}}{\epsilon_{0}} \cdot \frac{e^{-\frac{1}{2}l_{B}^{2}\mathbf{q}_{\perp}^{2}}}{\kappa \mathbf{q}^{2} + q_{\mathrm{TF}}^{2}(q_z) e^{-\frac{1}{2}l_{B}^{2}\mathbf{q}_{\perp}^{2}}}
\end{equation}
where $q_\mathrm{TF}(q_z)$ is the effective Thomas-Fermi wavevector
\begin{equation}
q_\mathrm{TF}^{2}(q_z)=\frac{e^2 M_{0}\ln\left|\frac{\sin ak_{c}+\sin\frac{aq_{z}}{2}}{\sin ak_{c}-\sin\frac{aq_{z}}{2}}\right|}{2\epsilon_{0}\pi^{2}l_{B}^{2}\left(M_{0}^{2}-C_{0}^{2}\right)a\sin\frac{aq_{z}}{2}}
\end{equation}
We have checked this approximation by comparing it with full self-consistent calculations,
 where $W^\mathrm{S}$ is calculated from $G$ self-consistently, and find that the correction
 on results is very small.

With the above approximation, the Dyson's equation $ ({\hat G}^{0-1}-{\hat \Sigma}){\hat G} = \mathbb{I} $
 (Fig. (\ref{fig:COHSEX}a)) and the equations (\ref{eq:Sigma-H}) and (\ref{eq:Sigma-E}) set up a self-consistent loop to determine
 the possible symmetry breaking phases at zero temperature by assuming different non-diagonal matrix elements in the self energy matrix.
For convenience, we define the order parameter as
 $\Delta_{an,bm}(k_x,k_z)=\langle \psi^\dagger_{an}(k_x,k_z)\psi_{bm}(k_x,k_z) \rangle$,
 whose non-diagonal elements in the band index $a,b$ and  sub-BZ index $n,m$ denote
 the appearance of the nematic phase and the CDW phase respectively.

\section{CDW phase}
CDW phase acquires its instability
from the Fermi surface nesting in
the quasi one dimensional band structure (Fig. (\ref{fig:channel})).
At first sight, it seems that the CDW should occur simultaneously
at $Q=2k_{c}+2k_{F}$ and $Q=2k_{c}-2k_{F}$ channels for conduction
and valence bands respectively. However, the interband Hartree energy
can lock the CDWs in different bands to same $Q=2k_{c}$
at least for low enough carrier density. This conclusion can
be reached by simply comparing the energy difference between the CDW
phases with $Q=2k_{c}\pm2k_{F}$ and $Q=2k_{c}$.
According to Eq. (\ref{eq:Sigma-H}), the $Q=2k_{c}$ phase gains an extra
 interband Hartree energy of $\sim W \left(Q\right)
 \Re e(\Delta_{c0,c-1}^\mathrm{CDW}\Delta_{v0,v-1}^\mathrm{CDW*})$,
 which reaches a negative constant as $k_F$ approaching zero if
 $\Delta_{c0,c-1}^\mathrm{CDW}=-\Delta_{v0,v-1}^\mathrm{CDW}$.
While the kinetic energy and exchange energy (Eq. (\ref{eq:Sigma-E}))
 difference between the $Q=2k_{c}$ and the $Q=2k_{c}\pm2k_F$ phases will vanish
 with $k_F$ approaching zero.
Therefore as long as $k_F$ is small enough, the CDW phase with $Q=2k_{c}$ for
both bands will be stabilized.

\begin{figure}
\begin{centering}
\includegraphics[width=1\linewidth]{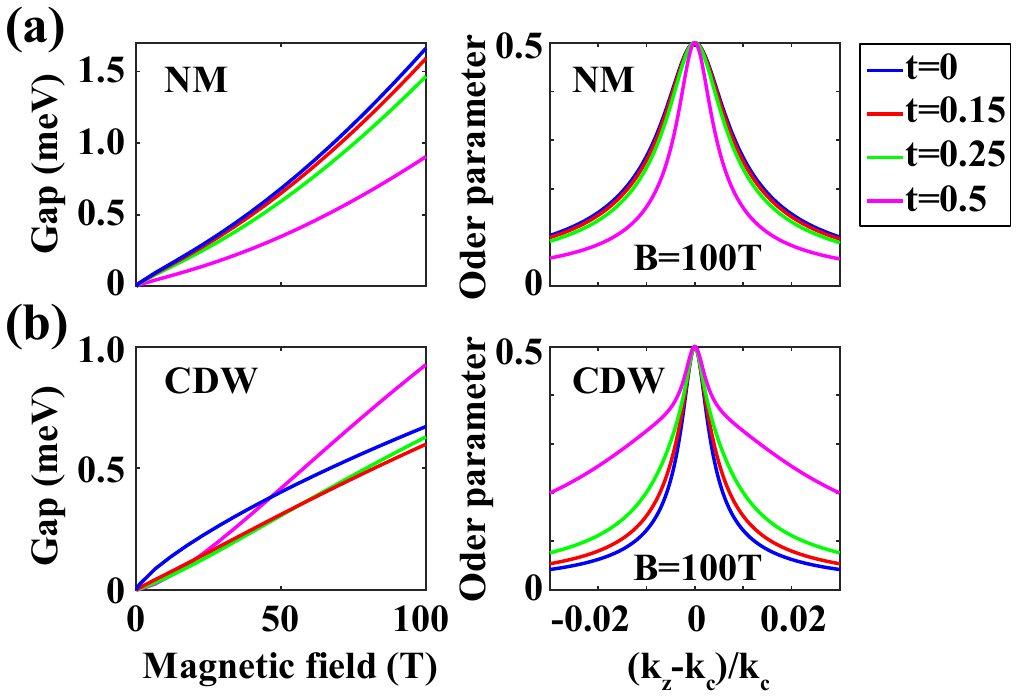}
\par\end{centering}

\protect\caption{\label{fig:gap}
In the left panel, the energy gaps of the polarized nematic and CDW phases
 at zero doping and a few tilting ratios are plotted as functions of magnetic field.
In the right panel, the corresponding order parameters of the two phases at $B=100\mathrm{T}$, i.e. $|\Delta_{v;c}^\mathrm{NM}(k_z)|$ and
 $\sqrt{\sum_a|\Delta_{a,-1;a,0}^\mathrm{CDW}|^2}$,
 are plotted around the DPs.
}
\end{figure}

The numerical calculation is performed with the initial condition
 $\Delta_{an,bm}^\mathrm{CDW}(k_x,k_z) =
 \delta_{n,m+1}\eta_{a,b}(k_z) + \delta_{n+1,m}\eta_{b,a}^*(k_z)$,
 where $\eta(k_z)$ is a random matrix.
The parameters are set as
 $\kappa_{0}=5$, $a_{0}=9.66\mathrm{\mathring{A}}$,
 $M_{0}a_{0}=2.3\mathrm{eV\cdot\mathrm{\mathring{A}}}$,
 $\hbar v=2.0\mathrm{eV\cdot\mathring{A}}$
 and $D_{c}=4$, which give the same Dirac velocity for Na$_3$Bi with the first principle results. \cite{wang_dirac_2012}
We set $C_0$ as $C_0=-tM_0$, where $t\in[0,1)$ is the tilting ratio
 describing how much the bands are tilted.
In Fig. (\ref{fig:gap}b), we plot the band gaps and order parameters
 at various tilting ratios and magnetic fields.
It shows that the tilting can significantly enlarge the CDW order,
 which is a direct consequence of saving the kinetic energy,
 as sketched in Fig. (\ref{fig:channel}b).


\begin{figure}
\begin{centering}
\includegraphics[width=0.9\linewidth]{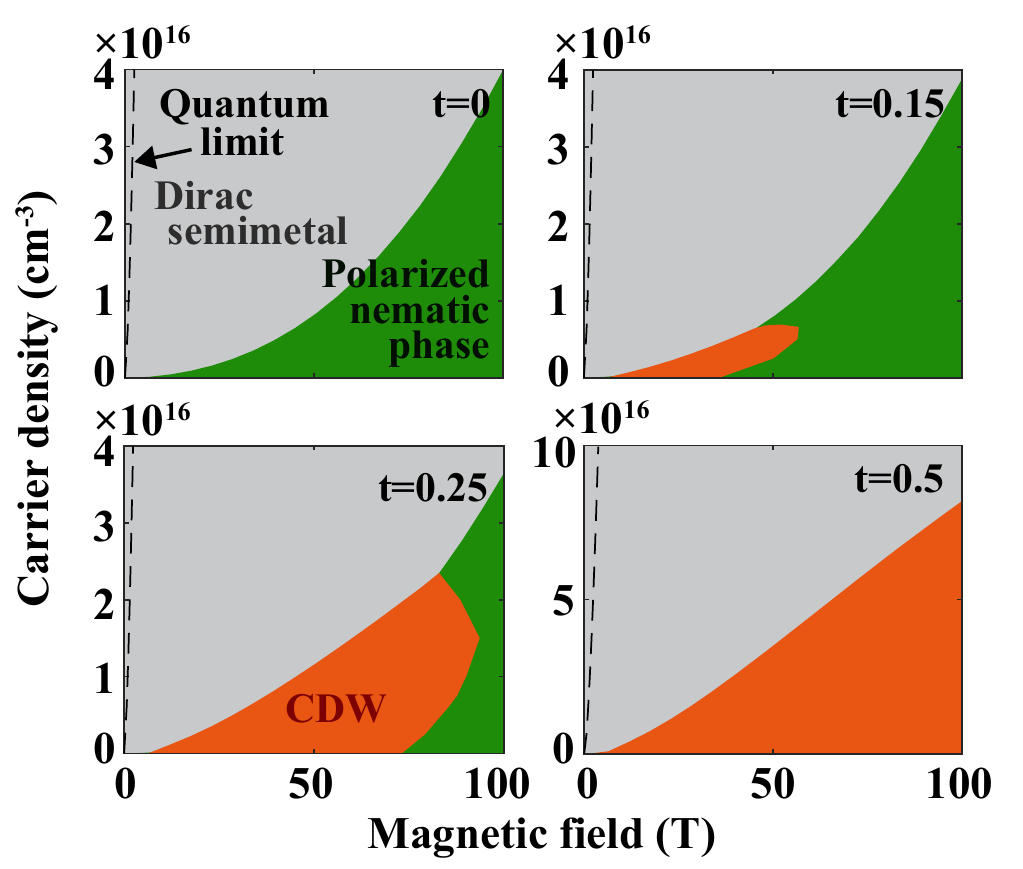}
\par\end{centering}

\protect\caption{\label{fig:PD}
The phase diagram in density-field parameter space at a few
 tilting ratios.
Here the critical field to achieve the quantum limit are indicated by dashed lines,
 and, the Dirac semimetal, polarized nematic, and CDW phases are represented
 by gray, green, and orange areas, respectively.
It shows that the polarized nematic phase is more favoured in untilted bands,
 while the CDW phase is more favoured in tilted bands.
 }
\end{figure}

\section{Nematic phase}

As shown in Fig. (\ref{fig:channel}), if the chemical potential
 is  close enough to the DPs a rotation broken phase, i.e. the nematic phase, can be stabilized.
Since the nematic phase doesn't break the translational symmetry, its order parameter can be expressed
 in the full BZ as $\Delta_{a,b}^\mathrm{NM}(k_x,k_z)=\delta_{\bar{a}b} \eta(k_z)$,
 where $-\pi\le k_z<\pi$, and $\bar{a}=v (c)$ for $a=c(v)$.
Two different types of $\eta$ can be got: odd or even with respect
to $k_{z}$.
According to the definition of LB wave function \cite{sup}, inversion operator acts
 on it as
 \begin{equation}
 \mathcal{P} \hat{\psi}_{k_{x},k_{z},c}^{\dagger} \hat{\psi}_{k_{x},k_{z},v} \mathcal{P}^{-1}
  =  -\hat{\psi}_{-k_{x},-k_{z},c}^{\dagger} \hat{\psi}_{-k_{x},-k_{z},v}
 \end{equation}
, thus the even and odd $\eta$ will respectively break and
 maintain the inversion symmetry.
As will be discussed in the next paragraph, the inversion broken phase,
 which will be referred as the polarized nematic phase in the following,
 is always more favored.

The band gaps and order parameters of the polarized nematic phase, and the phase diagram
consisting of the phases mentioned above
 are calculated with the same parameters used for the CDW phase
 and shown in Fig. (\ref{fig:gap}a) and (\ref{fig:PD}) respectively,
 which indicates that the polarized nematic phase is more favoured in untilted bands
 while the CDW phase is more favoured in tilted bands.
This can be understood as a result of competition between kinetic energy
 and interaction energy.
One one hand, as will be explained latter, the polarized nematic phase has
 a lower interaction energy; on the other hand, as shown in Fig. (\ref{fig:channel}b),
 the tilting will significantly lower the kinetic energy cost in the CDW phase.
Therefore, as shown in Fig. (\ref{fig:PD}), the area of the polarized nematic phase
 in phase diagram will shrink and eventually vanish with
 an increasing tilting.
Now let us explain why the polarized nematic phase has a lower interaction energy.
Since its Hartree energy reaches zero, i.e. the minimum,
 we only need to compare the exchange energies.
Eq. (\ref{eq:Sigma-E}) suggests that the exchange
 energy in the CDW phase is approximately
 $- W^\mathrm{S}(\mathbf{q}_\perp,0)
 |\Delta_{a,0;a,-1}^\mathrm{CDW}|^2$.
While the exchange energy in the nematic phase consists of three parts:
 two intravalley parts
 $ -  \frac{1}{2}W^\mathrm{S}(\mathbf{q}_\perp,0)
  |\Delta_{a\bar{a}}^\mathrm{NM}(\pm k_c)|^2$,
 which equal to the CDW one;
 and an intervalley part
   $- W^\mathrm{S}(\mathbf{q}_\perp,Q)
   \Re e( \Delta_{a\bar{a}}^\mathrm{NM}(k_c) \Delta_{a\bar{a}}^\mathrm{NM*} (-k_c) )$,
  which is negative in the polarized nematic phase ($\Delta_{a\bar{a}}^\mathrm{NM}(k_c)=\Delta_{a\bar{a}}^\mathrm{NM}(-k_c)$).
Here we have omitted the summation and integral symbols for brevity.
Thus we conclude that the polarized nematic order has lower interaction energy
 than the inversion symmetric nematic order and the CDW order.

Another aspect to understand this nematic order is to view it as a ``pairing order" 
 between electrons in the conduction band and holes in the valence band, which is the 
 “exciton condensation” state in the mean field level. \cite{excitonic_insulator}
Formally we can rewrite the creation operators of electrons and holes as 
 $\psi_{k_x,k_z,c}^\dagger = \psi_{k_x,k_z}^{\mathrm{e}\dagger}$,
 $\psi_{k_x,k_z,v} = \psi_{-k_x,-k_z}^{\mathrm{h}\dagger}$,
 and rewrite the order parameter as a pairing order
 $\langle \psi_{k_x,k_z}^{\mathrm{e}\dagger}  \psi_{-k_x,-k_z}^{\mathrm{h}\dagger} \rangle$.
Then the exchange interaction turns into an effective attractive interaction between the electrons in the
conduction band and holes in the valence band.
 And our mean field theory is equivalent to the BCS theory for superconductivity.
Since the system is three dimensional, the quantum fluctuation and disorder can not 
 suppress the phase coherence completely and such a transition 
 can survive even beyond the mean field approximation. 

\begin{figure}
\begin{centering}
\includegraphics[width=1\linewidth]{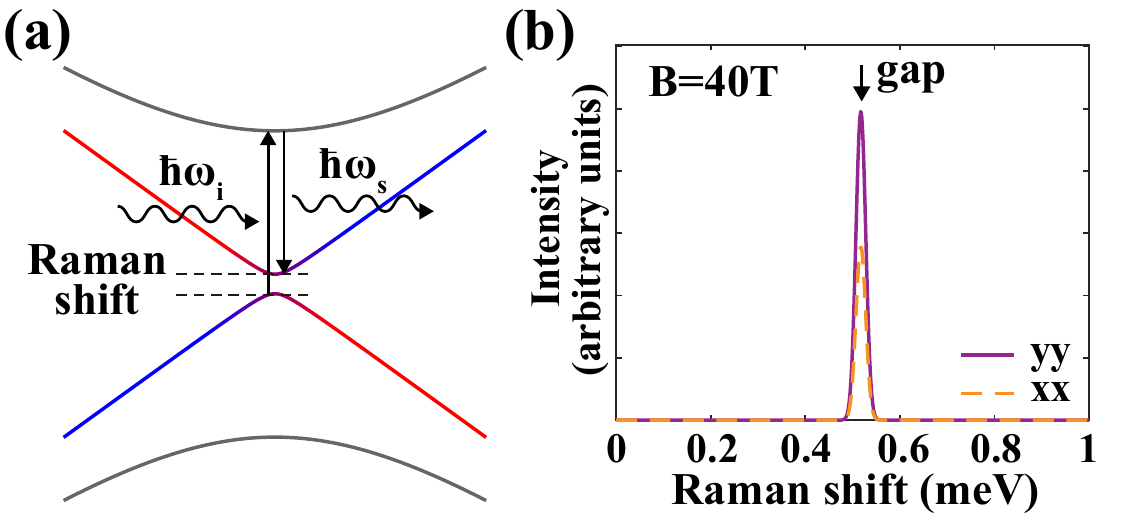}
\par\end{centering}

\protect\caption{\label{fig:Raman}
Raman scattering characterizing the nematic phase.
In (a) the scattering process that
 reveals the rotation symmetry breaking is sketched.
In (b) the numerical differential scattering section at zero doping
and $B=40\mathrm{T}$ is shown.}

\end{figure}

\section{Experimental aspects}
The most direct consequence of both the nematic and CDW phase transitions
 is the opening of an energy gap between the zeroth LBs, which
 can  be observed easily through the transport measurement.
For the CDW phase, since the order wave-vector given by the distance between two DPs
 is in general incommensurate with the lattice, the corresponding Goldstone
 mode, i.e. the so called sliding mode, will contribute to an electric field
 dependent conductivity along the wave-vector direction due to the depinning effect. \cite{SlidingCDW}
While, for the nematic phase, an anisotropic resistance in the $xy$ plane is expected
 due to the rotational symmetry breaking. Since the original rotational symmetry is discrete, the
 corresponding Goldstone mode in the nematic phase will be gapped and can de detected by neutron
 scattering experiments.

Another evidence for the nematic phase should be the anisotropy in the
 inelastic light scattering shown in Fig. (\ref{fig:Raman}a),
 where a strongly anisotropic scattering section with a Raman shift of the band gap
 will be observed, since both the initial and final states are rotation broken.
To verify this, we apply a numerical study of the Raman scattering section with the formula
$\frac{\partial^{2}\sigma}{\partial\Omega\partial\omega_{s}} \propto
\sum_{F}\left|M_{F,G}\right|^{2}\delta\left(E_{F}-E_{I}-\hbar\Omega\right)
$
where $\Omega=\omega_{i}-\omega_{s}$ is the Raman shift and $M_{F,I}$
is the light scattering matrix element \cite{devereaux_inelastic_2007}:
\begin{align}
M_{F,I} & =\mathbf{e}_{i}\cdot\mathbf{e}_{s}\left\langle F\right|\hat{\rho}\left|G\right\rangle \nonumber \\
+ & \frac{1}{m}\sum_{J}\left[\frac{\left\langle F\right|\hat{\pi}^{s}\left|J\right\rangle \left\langle J\right|\hat{\pi}^{i}\left|G\right\rangle }{E_{G}-E_{J}+\omega_{i}}+\frac{\left\langle F\right|\hat{\pi}^{i}\left|J\right\rangle \left\langle J\right|\hat{\pi}^{s}\left|G\right\rangle }{E_{G}-E_{J}-\omega_{s}}\right]
\end{align}
Here $\left|G\right\rangle $, $\left|J\right\rangle $ and $\left|F\right\rangle $
 represent the initial (ground), intermediate and final many body states
 having energies $E_{G,J,F}$, respectively.
$i,s$ represent the polarization direction of the initial and scattered photons,
 respectively.
And, $\hat{\rho}$, $\hat{\pi}$ are the density and velocity operators
 in second quantization form, respectively.
Results at zero doping and field $B=40\mathrm{T}$
 is shown in Fig. (\ref{fig:Raman}b), where the large splitting
 in the $xx$ and $yy$ polarized light measurements indicates the rotation
 symmetry breaking.

\section{Summary}
In summary, we have systematically studied the instabilities of Dirac semimetal
 phase in the quantum limit due to the Coulomb interaction.
The high LB electrons far away from the Fermi level are considered as a background
 to screen the interaction by an effective dielectric constant in the long wavelength limit.
All the possible instabilities on the zeroth LBs,
 i.e. the inter/intra-valley and inter/intra-band channels,
 are treated within the so called COHSEX method.
By numerical calculations, we have shown that
 a polarized nematic phase breaking both the rotational and inversion symmetry
 and a CDW phase breaking translational symmetry will be stabilized
 depending on the strength of the tilting terms for the Dirac cones.
Relevant experiments, including transport and Raman scattering,
 are also proposed to verify the existence of such phases.
Further theoretical studies on the physical properties like magneto-transport
 in these exotic phases are also strongly encouraged.

\section*{Acknowledgement}
We acknowledge the supports from National Natural Science Foundation of China,
the National 973 program of China (Grant No. 2013CB921700) and the ``Strategic Priority Research Program (B)" of the Chinese Academy of Sciences (Grant No. XDB07020100).

\appendix

\section{Solution of the free Hamiltonian}\label{sec:solution}
The eigenenergies and eigenstates of our model Hamiltonian can be explicitly derived as
\cite{abrikosov_quantum_1998}
\begin{equation}
\epsilon_{k_{z}s\alpha}=
\begin{cases}
C(k_z)+\sqrt{M^{2}\left(k_{z}\right)+2\hbar^{2}v^{2}l_{B}^{-2}\left|\alpha\right|} & \alpha>0\\
C(k_z)-sM\left(k_{z}\right) & \alpha=0\\
C(k_z)-\sqrt{M^{2}\left(k_{z}\right)+2\hbar^{2}v^{2}l_{B}^{-2}\left|\alpha\right|} & \alpha<0
\end{cases}
\end{equation}
\begin{align}
\hat{\psi}_{k_{x}k_{z}s\alpha}^{\dagger} & =\frac{1}{\sqrt{l_{z}l_{x}}}\sum_{j_z}
 \delta_{s,\mathrm{sign}(j_z)} C_{j_z}^{k_{z}s\alpha}  \nonumber \\
  \times & \int d^{3} \mathbf{r} e^{i\left(k_{z}z+k_{x}x\right)}
  H_{\alpha^\prime} \left(\frac{y}{l_B}-l_B k_{x}\right) \hat{\psi}_{j_z}^{\dagger}\left(\mathbf{r}\right)\label{eq:LBs}
\end{align}
, where $s=1(-1)$ represents the left up (right down) block in the Hamiltonian,
  $\alpha=$$0,\pm1,\cdots$ is the LB index,
  $j_z = \frac{3}{2},\frac{1}{2},-\frac{1}{2},-\frac{3}{2}$ is the
  k$\cdot$p basis index,
  and $H_{\alpha^\prime}$
 is the $\alpha^\prime$-th order one-dimensional harmonic oscillator.
Here $\alpha^\prime$ is defined as $\alpha^\prime=|\alpha|+s-j_z-\frac{1}{2}$,
 which equals to $|\alpha|-1,|\alpha|,|\alpha|-1,|\alpha|$
 for $j_z = \frac{3}{2},\frac{1}{2},-\frac{1}{2},-\frac{3}{2}$
 respectively, and $\alpha^\prime = -1$ terms should be omitted.
The coefficient $C_{j_z}^{k_z s \alpha}$ is defined as
\begin{align}
C_\frac{3}{2}^{k_z,+1,\alpha}&=\cos\frac{\theta}{2} \qquad
C_\frac{1}{2}^{k_z,+1,\alpha}=\sin\frac{\theta}{2} \nonumber \\
C_{-\frac{1}{2}}^{k_z,-1,\alpha}&=-\sin\frac{\theta}{2} \qquad
C_{-\frac{3}{2}}^{k_z,-1,\alpha}=\cos\frac{\theta}{2}
\end{align}
for $\alpha>0$,
\begin{align}
C_\frac{3}{2}^{k_z,+1,\alpha}&=-\sin\frac{\theta}{2} \qquad
C_\frac{1}{2}^{k_z,+1,\alpha}=\cos\frac{\theta}{2} \nonumber \\
C_{-\frac{1}{2}}^{k_z,-1,\alpha}&=\cos\frac{\theta}{2} \qquad
C_{-\frac{3}{2}}^{k_z,-1,\alpha}=\sin\frac{\theta}{2}
\end{align}
for $\alpha<0$, and
\begin{align}
C_\frac{3}{2}^{k_z,+1,\alpha}&=0 \qquad
C_\frac{1}{2}^{k_z,+1,\alpha}=1 \nonumber \\
C_{-\frac{1}{2}}^{k_z,-1,\alpha}&=0 \qquad
C_{-\frac{3}{2}}^{k_z,-1,\alpha}=1
\end{align}
for $\alpha=0$, respectively, where the auxiliary angle $\theta$ is set by
 $\theta=\arctan \frac{v\hbar\sqrt{2|\alpha|}}{M(k_z)l_B}$
 ($0\le \theta<\pi$).

The conduction and valence bands in the paper are the $\alpha=0,s=1$
 and $\alpha=0,s=-1$ bands here.

\section{Effective interaction on the zeroth LBs} \label{sec:RPA}

In this section, we will derive the effective interaction on the zeroth
LBs by tracing out the high LBs in RPA.
The long range Coulomb interaction can be written as
\begin{align}
\hat{H}_{\mathrm{int}} & =\frac{1}{2}\sum_{j_z j_z^\prime}\int d^{3} \mathbf{r}
 \int d^{3}\mathbf{r}^{\prime} \frac{e^{2}}{4\pi\epsilon_{0}\kappa_{0}\left|\mathbf{r}-\mathbf{r}^\prime\right|}
 \nonumber \\
 & \times
 \hat{\psi}_{j_z }^{\dagger}\left(\mathbf{r}\right)
 \hat{\psi}_{j_z^\prime}^{\dagger}\left(\mathbf{r}^{\prime}\right)
  \hat{\psi}_{j_z^\prime}\left(\mathbf{r}^{\prime}\right) \hat{\psi}_{j_z}\left(\mathbf{r}\right)\label{eq:Hint_0}
\end{align}
where $\kappa_{0}$ is the dielectric constant contributed by the
 core electron states.
By a representation transformation, the interaction can be written on the LB bases
\begin{align}
\hat{H}_{\mathrm{int}} & = \frac{1}{2\Omega} \sum_{\mathbf{q}\neq0} \sum_{k_x k_z} \sum_{p_x p_z}
 \sum_{ss^\prime} \sum_{\alpha\beta\alpha^{\prime}\beta^{\prime}} e^{il_{B}^{2}q_{y}\left(k_{x}-p_{x}\right)}\nonumber \\
 & \times U_{k_{z}s\alpha\alpha^{\prime},p_{z}s^\prime\beta \beta^{\prime}} \left(\mathbf{q}\right) \hat{\psi}_{k_{x}+q_{x},k_{z}+q_{z},s\alpha^{\prime}}^{\dagger} \hat{\psi}_{p_{x},p_{z},s^\prime\beta}^{\dagger}\nonumber \\
 & \times\hat{\psi}_{p_{x}+q_{x},p_{z}+q_{z},s^\prime\beta^{\prime}}
 \hat{\psi}_{k_{x},k_{z},s\alpha}
\end{align}
where
\begin{align}
U_{k_{z}s\alpha\alpha^{\prime},p_{z}s^\prime\beta\beta^{\prime}} \left(\mathbf{q}\right) &
 =\frac{e^{2}}{\epsilon_{0}\kappa_{0}\mathbf{q}^{2}}
 e^{-\frac{1}{2}l_{B}^{2}\mathbf{q}_\perp^2}\nonumber \\
 & \times \Lambda_{k_{z}s\alpha\alpha^{\prime}}^{*} \left(\mathbf{q}\right)
 \Lambda_{p_{z}s^\prime\beta\beta^{\prime}} \left(\mathbf{q}\right)
\end{align}
\begin{align}
&\Lambda_{p_{z}s^\prime\beta\beta^{\prime}}\left(\mathbf{q}\right)  =\sum_{j_z}
 \delta_{s^\prime,\mathrm{sign}(j_z)} C_{j_z}^{p_{z}s^\prime\beta*}
 C_{j_z}^{p_{z}+q_{z},s^\prime\beta^{\prime}}\nonumber \\
&\quad \times
 F_{\left|\beta\right|+s^\prime-j_z-\frac{1}{2},\left|\beta^{\prime}\right|+s^\prime-j_z-\frac{1}{2}}
 \left(\frac{l_{B}\left(q_{x}-iq_{y}\right)}{\sqrt{2}}\right)
\end{align}
Here $F_{\alpha,\beta}\left(\xi\right)$ is the well known form factor
of Landau levels, which is defined as
\begin{equation}
F_{\alpha,\beta}\left(\xi\right)=\sqrt{\frac{\beta!}{\alpha!}}\xi^{\alpha-\beta}L_{\beta}^{\left(\alpha-\beta\right)}\left(\left|\xi\right|^{2}\right)
\end{equation}
for $\alpha\ge\beta$ and $F_{\alpha,\beta}\left(\xi\right)=F_{\beta,\alpha}^{*}\left(-\xi\right)$
for $\alpha\le\beta$ , and $L_{\beta}^{\left(\alpha-\beta\right)}$
is the Laguerre polynomial \cite{macdonald_introduction_1994,cahill_ordered_1969}.

\begin{figure}
\begin{centering}
\includegraphics[width=1\linewidth]{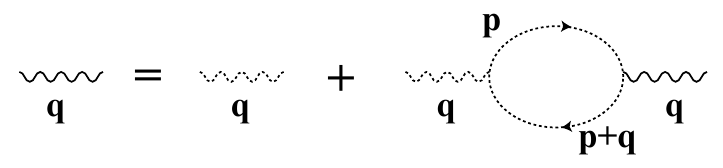}
\par\end{centering}

\protect\caption{\label{fig:Ueff}
RPA diagrams of the effective interaction on the zeroth LBs.
The solid wavy line represents the effective interaction,
 while the dashed wavy line represents the bare interaction.
The dashed straight line represents the Green's functions in high LBs.}
\end{figure}

In the Feynman's diagram representation, the RPA effective interaction on the zeroth
 LBs can be interpreted as the ``dressed'' interaction, that has
 been inserted with bubble diagrams concerning high
 LBs (Fig. (\ref{fig:Ueff})).
Thus the static effective interaction satisfy
\begin{equation}
W\left(\mathbf{q}\right) = U\left(\mathbf{q}\right) + U\left(\mathbf{q}\right) \chi^{0>} \left(0q_{x}q_{z}\right) W\left(\mathbf{q}\right)
\end{equation}
, where the matrix subscripts are omitted.
$\chi^{0>}$ is the bare susceptibility of high LBs:
\begin{align}
 & \chi_{k_{x}k_{z}s\alpha\beta, k_{x}^{\prime} k_{z}^{\prime} s^{\prime} \alpha^{\prime} \beta^{\prime}}^{0>}
 \left(\omega q_{x}q_{z}\right) = \delta_{k_{x}k_{x}^{\prime}} \delta_{k_{z}k_{z}^{\prime}} \delta_{ss^{\prime}} \delta_{\alpha\alpha^{\prime}} \delta_{\beta\beta^{\prime}}\nonumber \\
 & \times\frac{1}{\Omega}\begin{cases}
\frac{n_{F}\left(\epsilon_{k_{z}s\alpha}-\mu\right)-n_{F}\left(\epsilon_{k_{z}+q_{z}s\beta}-\mu\right)} {\omega+\epsilon_{k_{z}s\alpha}-\epsilon_{k_{z}+q_{z}s\beta}} & \alpha,\beta\neq0\\
0 & \text{else}
\end{cases}
\end{align}
Therefore the effective interaction can be derived as
\begin{align}
W_{k_{z}s,p_{z}s^\prime}\left(\mathbf{q}\right) & = \left[ U \left( \mathbf{q} \right) \left(1-\frac{\chi^{0>}U\left(\mathbf{q}\right)}{\Omega}\right)^{-1}\right]_{k_{z}s00;p_{z}s^\prime00}\nonumber \\
 & =\frac{e^{2}}{\epsilon_{0}\kappa\left(\mathbf{q}\right)\mathbf{q}^{2}}e^{-\frac{1}{4}l_{B}^{2}\mathbf{q}_{\perp}^{2}}
\end{align}
where
\begin{align}
\kappa\left(\mathbf{q}\right) & =\kappa_{0}-\frac{e{}^{2}}{2\pi\epsilon_{0}l_{B}^{2} \mathbf{q}^{2}}e^{-\frac{1}{2}l_{B}^{2} \mathbf{q}_\perp^2} \sum_{s\alpha\beta}\nonumber \\
 & \times \int \frac{dk_{z}^{\prime}}{2\pi}
  \Lambda_{k_{z}^{\prime}s\alpha\beta}^{*}\left(\mathbf{q}\right) \chi_{k_{z}^{\prime}s\alpha\beta}^{0>}\left(q_{z}\right)
 \Lambda_{k_{z}^{\prime}s\alpha\beta}\left(\mathbf{q}\right)\label{eq:kap1-def}
\end{align}
is the effective dielectric function.
For brevity, here we use $\chi_{k_{z}s\alpha\beta}^{0>}\left(\omega q_{z}\right)$ to represent
  the diagonal elements of $\chi^{0>}$.
As $W_{k_{z}s,p_{z}s^\prime}\left(\mathbf{q}\right)$
 does not depend on its subscripts, we will denote it as $W\left(\mathbf{q}\right)$
 in the paper.

\section{Long wave behavior of the effective interaction} \label{sec:kap}

\begin{figure}
\begin{centering}
\includegraphics[width=0.85\linewidth]{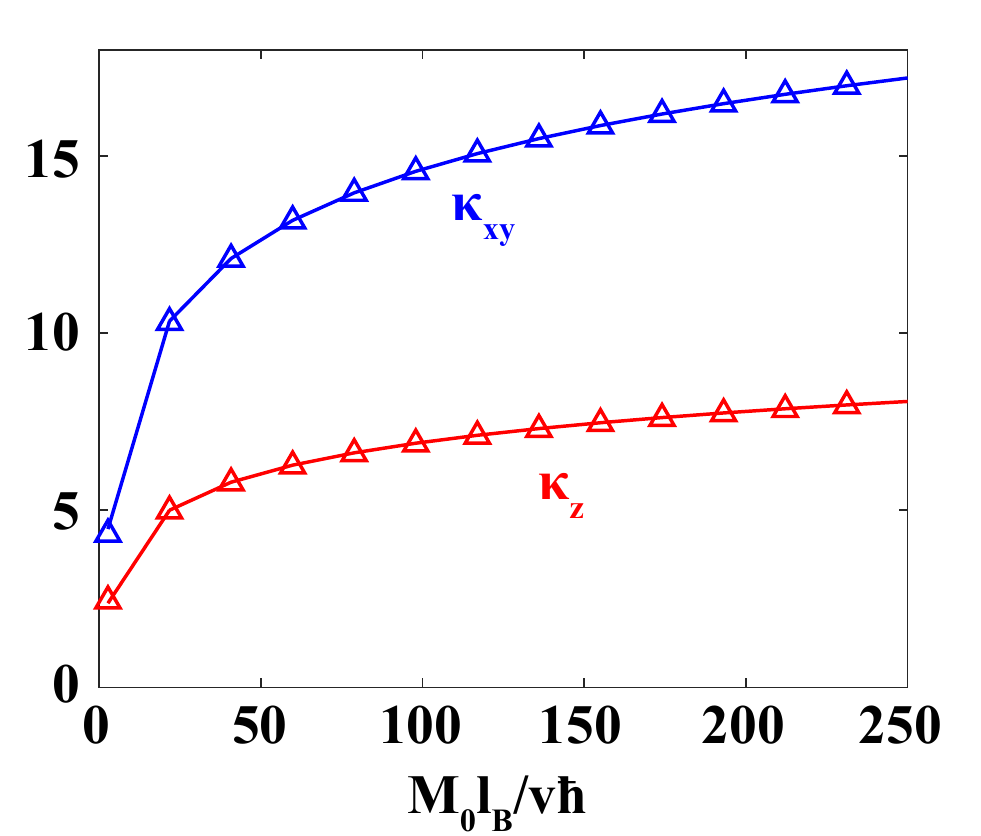}
\par\end{centering}

\protect\caption{\label{fig:kap}
The dielectric constants calculated from definition equations ( Eq. (\ref{eq:kap-para}) and (\ref{eq:kap-perp}) )
 and simplified equations ( Eq. (\ref{eq:kap-para-a}) and (\ref{eq:kap-perp-a}) )
 are plotted by triangles and lines, respectively.
It shows that the simplified equations give a very good approximation.
}
\end{figure}

In this section, we intend to get a more explicit expression of the
dielectric function in the long wavelength limit.
Expand $\Lambda_{k_{z}^{\prime}a\alpha\beta}^{*}\left(\mathbf{q}\right)\chi_{k_{z}^{\prime}a\alpha\beta}^{0>}\left(q_{z}\right)\Lambda_{k_{z}^{\prime}a\alpha\beta}\left(\mathbf{q}\right)$
to second order of $\mathbf{q}$, we have
\begin{align}
 & \sum_{\alpha\beta}\Lambda_{k_{z}^{\prime}a\alpha\beta}^{*}\left(\mathbf{q}\right)\chi_{k_{z}^{\prime}a\alpha\beta}^{0>}\left(q_{z}\right)\Lambda_{k_{z}^{\prime}a\alpha\beta}\left(\mathbf{q}\right)\nonumber \\
\approx & \sum_{\alpha}\chi_{k_{z}^{\prime}a\alpha,-\alpha}^{0>}\left(q_{z}\right)\left[\frac{1}{4}\left(\frac{\partial\theta_{\alpha}}{\partial k_{z}}q_{z}\right)^{2}+\frac{l_{B}^{2}\left(q_{x}^{2}+q_{y}^{2}\right)\sin^{2}\theta_{\alpha}}{16\left|\alpha\right|}\right]
\end{align}
Substitute the definition of the auxiliary angle $\theta_{\alpha}$
 in, we get
\begin{align}
\kappa\left(\mathbf{q}\right) & \approx \kappa_{0}
 +\kappa_{z}\cos^{2}\left\langle \mathbf{q},\mathbf{B}\right\rangle
 +\kappa_{xy}\sin^{2}\left\langle \mathbf{q},\mathbf{B}\right\rangle
\end{align}
where
\begin{equation}\label{eq:kap-para}
\kappa_{z} = \frac{e{}^{2}}{8\pi^{2}\epsilon_{0}l_{B}^{2}} \sum_{\alpha=1}^{\infty} \int \frac{2v^{2}\hbar^{2}l_{B}^{-2}\alpha\left(\frac{\partial M\left(k_{z}\right)}{\partial k_{z}}\right)^{2}dk_{z}^{\prime}}{\left(M^{2}\left(k_{z}\right)+2v^{2}\hbar^{2}l_{B}^{-2}\alpha\right)^{\frac{5}{2}}}
\end{equation}
\begin{equation}\label{eq:kap-perp}
\kappa_{xy} = \frac{e{}^{2}}{8\pi^{2} \epsilon_{0}l_{B}^{2}} \sum_{\alpha=1}^{\infty} \int \frac{2v^{2}\hbar^{2}dk_{z}^{\prime}} {4\left(M^{2}\left(k_{z}\right)+2v^{2}\hbar^{2}l_{B}^{-2}\alpha\right)^{\frac{3}{2}}}
\end{equation}
and $\left\langle \mathbf{q},\mathbf{B}\right\rangle $ is the angle
between $\mathbf{q}$ and the magnetic field. Eq. (\ref{eq:kap-para})-(\ref{eq:kap-perp})
may be simplified further. Firstly, as the main contribution in the
$k_{z}^{\prime}$ integral comes from small $M\left(k_{z}\right)$,
we can expand $M\left(k_{z}\right)$ to linear order of $k_{z}$ around
each DP. Secondly, the limit $v\hbar l_{B}^{-1}\ll M_{0}$ is assumed
such that the Landau level splitting is significantly smaller than
the bandwidth and so the summation over $\alpha$ can be approximated
by integral. Therefore, we achieve the following formula
\begin{equation}\label{eq:kap-para-a}
\kappa_{z} \approx \frac{e{}^{2}u}{3\pi^{2} \epsilon_{0}v^{2}\hbar} \left(0.9+\ln\left(\frac{M_{0}l_{B}}{v\hbar}\right)\right)
\end{equation}
\begin{equation}\label{eq:kap-perp-a}
\kappa_{xy} \approx \frac{e{}^{2}}{4\pi^{2}\epsilon_{0}u\hbar} \left(0.6+\ln\left(\frac{M_{0}l_{B}}{v\hbar}\right)\right)
\end{equation}
where $u=\frac{1}{\hbar}M_{0}a_{0}\left|\sin\left(a_{0}k_{c}\right)\right|$
 is the Dirac velocity along $z$ direction, and the coefficients $0.9$
 and $0.6$ are got by fitting Eq. (\ref{eq:kap-para-a})-(\ref{eq:kap-perp-a})
 with Eq. (\ref{eq:kap-para})-(\ref{eq:kap-perp}) numerically.
Indeed, Eq. (\ref{eq:kap-para-a})-(\ref{eq:kap-perp-a})  give very good approximations
 for Eq. (\ref{eq:kap-para})-(\ref{eq:kap-perp}) in a quite wide range.
In Fig. (\ref{fig:kap}), we compare the two equations  with the parameters used
in the paper.

In the end, if we neglect the dependence of $\kappa$ on the direction of $\bf q$,
 a dielectric constant can be got by an average on the solid angle:
\begin{equation}
\kappa \approx\kappa_{0} + \frac{1}{3}\kappa_{z} + \frac{2}{3}\kappa_{xy}
\end{equation}

\bibliographystyle{apsrev4-1}
\bibliography{main}

\begin{thebibliography}{46}
\bibitem [{gfa()}]{gfactor}%
  \BibitemOpen
  \href@noop {} {\ }\bibinfo {note} {To be published.}\BibitemShut {Stop}%
\end{thebibliography}

\end{document}